\documentclass[letterpaper,twocolumn,nofootinbib,superscriptaddress,showpacs]{revtex4}
\pdfoutput=1 % To make arXiv use PDF 

\usepackage{amsmath}
\usepackage{bbm}
\usepackage{amsfonts}
\usepackage{graphicx}
\usepackage{verbatim}
\usepackage{times} % APS uses Times-Roman fonts
\usepackage{hyperref} % Setup follows
\hypersetup{%
pdftitle  = {Symmetric extension in two-way quantum key distribution},
pdfauthor = {Geir Ove Myhr, Joseph M. Renes, Andrew C. Doherty and Norbert L{\374}tkenhaus},
pdfkeywords = {quantum state, symmetric extension, quantum key distribution, QKD, extendible, extendable, extensible, extendibility, extendability, extensibility, Bell-diagonal, semidefinite programming},
colorlinks,
linkcolor=blue,
filecolor=blue,
urlcolor=black,
citecolor=blue
}

\newcommand{\ket}[1]{|#1\rangle}
\newcommand{\bra}[1]{\langle#1|}
\newcommand{\ketbra}[1]{\ket{#1}\bra{#1}}
\newcommand{\proj}[1]{\ket{#1}\bra{#1}}

\newcommand{\tr}{\mathrm{Tr}}

\newcommand{\etal}{\textit{et~al}}

\newcommand{\eprint}[2][]{\href{http://arxiv.org/abs/#2}{arXiv:#2}}
\newcommand{\doiref}[2]{\href{http://dx.doi.org/#1}{#2}}

\DeclareMathOperator{\rank}{rank}

\begin{document}

% The title is open for change
\title{Symmetric extension in two-way quantum key distribution}

% Ordered according to Joe's suggestion. Hope it's okay for everyone.

\author{Geir Ove \surname{Myhr}}

\email{gomyhr@iqc.ca}

\affiliation{
%Quantum Information Theory Group, 
Institut f\"{u}r Theoretische Physik I, and Max-Planck Institute for the Science of Light, 
Universit\"{a}t Erlangen-N\"{u}rnberg, Staudtstra{\ss}e 7/B2, 91058 Erlangen, Germany\\}

\affiliation{Department of Physics and Astronomy and Institute for Quantum Computing, University of Waterloo, 200 University Ave.~W., Waterloo Ontario N2L 3G1, Canada}

\author{Joseph M. Renes}
\affiliation{Institut f{\"u}r Angewandte Physik, Technische Universit{\"a}t Darmstadt, Hochschulstr.~4a, 64289 Darmstadt, Germany}

\author{Andrew C. Doherty}
\affiliation{School of Physical Sciences, The University of Queensland, Queensland 4072, Australia}

\author{Norbert L{\"u}tkenhaus}

% These affiliations must be exact copies of the ones above.
\affiliation{
%Quantum Information Theory Group, 
Institut f\"{u}r Theoretische Physik I, and Max-Planck Institute for the Science of Light, 
Universit\"{a}t Erlangen-N\"{u}rnberg, Staudtstra{\ss}e 7/B2, 91058 Erlangen, Germany\\}

\affiliation{Department of Physics and Astronomy and Institute for Quantum Computing, University of Waterloo, 200 University Ave.~W., Waterloo Ontario N2L 3G1, Canada}

\begin{abstract}
We introduce symmetric extensions of bipartite quantum states as a tool
for analyzing protocols that distill secret key from quantum correlations. 
Whether the correlations are coming from a prepare-and-measure 
quantum key distribution scheme or from an entanglement-based scheme,
the protocol has to produce effective states without a symmetric extension in order to succeed.
By formulating the symmetric extension problem as a semidefinite program,
we solve the problem for Bell-diagonal states.
Applying this result to the six-state and BB84 schemes,
we show that for the entangled states that cannot be distilled by current key distillation procedures,
the failure can be understood in terms of a failure to break a symmetric extension.
\end{abstract}

% PACS numbers: (http://www.aip.org/pacs/pacs08/pacs0800.html)
% 03.67.Dd: Quantum cryptography and communication security
% 03.67.Mn: Entanglement measures, witnesses, and other characterizations (symmetric extension goes under other characterizations)
\pacs{03.67.Dd, 03.67.Mn}

\maketitle

\section{Introduction}

An important quantity characterizing the performance of a quantum key distribution (QKD) 
scheme is the maximum
amount of channel noise which can be tolerated before the 
protocol fails to produce a secure key.

This threshold has clear implications for QKD as a potentially-realizable technology and not
just as a possibility inherent in the formalism of quantum mechanics. 
It also relates to an important issue of principle, namely, 
the connection between quantum mechanics and privacy. What aspects of quantum mechanics are responsible for
the possibility of key distribution and other cryptographic protocols?  
Determining the threshold, or at least bounding it, gives us insight into this issue. 
Particular properties of quantum states and channels which are sufficient for generating privacy in
some way lead to lower bounds, while identifying properties necessary for privacy leads to upper bounds. 

One such property is the symmetric extendibility of bipartite quantum states. 
Suppose that two honest parties Alice and Bob share a state $\rho^{AB}$ from which 
they would like to extract a secret key using one-way public communication from her side to his. 
This task is impossible should there exist a
tripartite state $\rho^{ABB'}$ such that the $AB'$ marginal state is identical to the $AB$ marginal
$\rho^{AB'}=\rho^{AB}$. 
If such a state exists it can always be chosen to be symmetric between $B$ and $B'$, 
i.e.,~the state is invariant when $B$ and $B'$ are swapped.
Such a tripartite state is called a symmetric extension of the original state, and 
the equality of the marginals means that the extra system $B'$ essentially functions as a copy of system $B$.
Assuming the worst-case scenario that an eavesdropper Eve holds $B'$, 
whatever process Bob uses to create an error free bit string  after
receiving the communication from Alice can also be performed by Eve, 
and thus the bit string cannot be private \cite{moroder06b}.

The question of symmetric extendibility is relevant in many areas of quantum information theory, 
from Bell-inequalities \cite{terhal03a} to quantum channel capacity \cite{nowakowski09a}. 
In QKD, the necessary condition of not having a symmetric extension has been translated into 
upper bounds on the key rate and threshold noise rate for one way procedures in~\cite{moroder06b}. 
A considerable advantage of this approach stems from the fact that the upper bounds are
determined without having to construct concrete eavesdropping attacks. 
Moreover, for systems described by Hilbert spaces of modest dimension, 
symmetric extensions can be efficiently constructed---when they exist---by means of semidefinite programming. 

In this paper, we consider the case of two-way communication and use the symmetric extension to 
derive attack-independent upper bounds for the BB84~\cite{bennett84a} and six-state~\cite{bruss98a} schemes. 
At first glance, symmetric extensions appear to be irrelevant to the problem, 
since the two-way nature of the communication creates an asymmetry between the honest and dishonest parties---Eve 
cannot pretend to be one of the honest parties. However, every two-way communication procedure 
consists of alternating rounds of one-way communication, 
which must eventually terminate if the protocol is to establish a secret key that can be used in other applications. 
The final step thus involves only one-way communication, 
and the question of symmetric extendibility again becomes relevant. 
From this point of view, it becomes clear that the goal of the two-way communication is to \emph{break} 
any existing symmetric extension of the input state. 

To avoid confusion, we distinguish between a QKD \emph{protocol}, 
a QKD \emph{scheme}, and the various \emph{procedures} such as advantage distillation, error correction, and privacy amplification.
By a QKD scheme, we mean the generation of correlated data by distributing quantum particles and measuring them. 
For simplicity we will also include the parameter estimation and sifting in the definition of the scheme. 
Including the sifting means that BB84 \cite{bennett84a} and SARG04 \cite{scarani04a} are different schemes, 
even though the signal states and measurements are the same. 
For a given scheme, the key distillation procedures following it can be chosen in different ways which give different thresholds.
We call the whole process a QKD protocol, so that at the end of the protocol the parties have a secret key that is ready for use.

Given a particular two-way procedure, then, the question of symmetric extendibility leads to an
upper bound on the noise threshold given by the noisiest state for which the procedure just fails to 
break the extension. 
Generally one must resort to finding an approximate symmetric extension by solving the semidefinite program numerically. 
For the well-known BB84~\cite{bennett84a} and six-state~\cite{bruss98a} schemes, however, 
the relevant states can be assumed without loss of generality to be diagonal in a basis of maximally entangled states: 
the so-called Bell basis. 
By making use of the symmetries of the Bell states, 
we can answer the question of symmetric extendibility by solving the semidefinite program analytically, 
which then leads to exact upper bounds for the tolerable error rates.
For the two-way procedure outlined by Chau~\cite{chau02a}, 
we show that the upper bound for the two schemes meets the lower bound given therein 
and that this procedure is optimal for a wide range of two-way procedures.  
This agrees with the results reported by Ac\'in \etal.~\cite{acin06a}, 
who based their upper bound on an explicit eavesdropping attack. 

Our results are organized as follows. 
In Sec.~\ref{sec:SE}, we examine in detail the role played by the symmetric extension in two-way QKD protocols.
In Sec.~\ref{sec:Bell} we review the formulation of the symmetric extension problem as a semidefinite program (SDP), simplify, and solve it for Bell-diagonal states and give 
an analytic expression for the boundary of extendible Bell-diagonal states. 
Section \ref{sec:qkd} describes Chau's two-way communication procedure and shows that above the known lower bound on the threshold, the procedure fails to break the symmetric extension and can therefore not lead to a secret key.
We also discuss variations of this procedure which turn out to be equivalent for distillability. 
In Sec.~\ref{sec:discussion} we sum up and discuss some open questions.

\section{Breaking Symmetric Extensions}
\label{sec:SE}

One goal of the portion of a QKD protocol involving two-way communication is to transform a state
having a symmetric extension into one which does not. 
In a prepare and measure (P\&M) scheme, there is never an actual bipartite entangled state, 
but any such scheme can be modeled as an entanglement-based scheme where Alice prepares an entangled state and sends half of it to Bob.
When Alice measures her half of the entangled state, 
this effectively prepares the other half in one of the signal states of the P\&M protocol~\cite{bennett92c}.
Eve may interfere with the transmitted signal in any manner of her choosing, 
so after making their respective measurements, 
Alice and Bob compare a portion of the data in order to determine---at least roughly---what particular quantum state $\rho^{AB}$ they share. 
This state is the starting point for our analysis and is any state obeying
\begin{align}
\label{eqn:pe}
p_{jk}=\tr\left[\rho^{AB}\left(A_j\otimes B_k\right)\right],
\end{align}
where $A_j$ and $B_k$ are the POVM elements of Alice's and Bob's respective measurements and $p_{jk}$ are the probabilities with which Alice and Bob obtain outcomes corresponding to 
$A_j$ and $B_k$, respectively. 
Any subsequent processing of the measurement data is then modeled as a coherent processing on the quantum states with any classical communication corresponding to measurement outcomes.
This allows us to track the effective state throughout the protocol.

Now we can investigate which two-way processing procedures can break symmetric extensions. 
By making a few assumptions on the form of the procedure, 
we can simplify the problem considerably. 
Assume that each round of one-way processing is performed on blocks with a finite number of systems, 
such that the output is considered as a single system in the next round. 
As we are not concerned with the rate of distillation, 
only whether the state is at all distillable or not,  
we are led to the following two simplifications.

First, we only need to concern ourselves with filtering operations---quantum operations defined by a single Kraus operator---which do not always succeed when applied to a state. 
There is a corresponding Kraus operator for failure, 
which makes the operation trace-preserving, 
but we discard the failure outcomes and therefore a filter is in general not trace-preserving.
If $\rho^{AB}$ is the state of the block before postprocessing, 
the unnormalized state after Bob applies the filter $K$ is 
$(\mathbbm{1} \otimes K) \rho^{AB} (\mathbbm{1} \otimes K)^\dagger$ 
and the filter satisfies $K^\dagger K \leq \mathbbm{1}$.
If an operation with more than one Kraus operator is able to break the symmetric extension --- 
that is $\sum_j (\mathbbm{1} \otimes K_j) \rho^{AB} (\mathbbm{1} \otimes K_j)^\dagger$ 
(where $\sum_j K_j^\dagger K_j \leq \mathbbm{1}$) has no symmetric extension --- 
then because of convexity at least one of the 
$(\mathbbm{1} \otimes K_j) \rho^{AB} (\mathbbm{1} \otimes K_j)^\dagger$ 
must be without symmetric extension, so the filter $K_j$ alone will break the extension.

Second, we can reduce the finite number of one-way rounds to only two, for the following reason.
Assume that the final round of communication is from Alice to Bob. 
Bob can start the procedure already at his last round 
by guessing ahead of time what Alice's messages related to that block would have been 
and perform the corresponding local operations. 
Usually this guess will be wrong and Alice will tell Bob to discard those blocks in the final round. 
For the tiny fraction of the blocks where Bob guessed correctly, 
Alice can proceed with her last round. 
This means that if the symmetric extension can be broken during a two-way procedure, 
it must also be possible to break it with a single filter on a block of copies of Bob's system.

\section{Extendibility of Bell-diagonal States}
\label{sec:Bell}

Bell-diagonal states are two-qubit states that are diagonal in the basis of maximally entangled states
$\ket{\Phi^\pm} = \frac{1}{\sqrt{2}} (\ket{00} \pm \ket{11})$, 
$\ket{\Psi^\pm} = \frac{1}{\sqrt{2}} (\ket{01} \pm \ket{10})$. 
Such states can be produced by sending 
half of the maximally entangled state $\ket{\Phi^+}$ through a Pauli channel
with error probabilities $p_x$, $p_y$, and $p_z$ for the $\sigma_x$, $\sigma_y$, and $\sigma_z$ errors respectively. 
This results in the state 
$\rho^{AB} = p_I \proj{\Phi^+} + p_x  \proj{\Psi^+} + p_y  \proj{\Psi^-} + p_z  \proj{\Phi^-}$, 
where $p_I = 1- p_x - p_y - p_z$.
For compactness, we will also denote this as 
$\rho^{AB} = \sum_j p_j \proj{\beta_j}$ 
where the index $j$ runs over the set $\{I,x,y,z\}$ or equivalently $\{0,1,2,3\}$.

In the six-state and BB84 QKD schemes considered here, the effective quantum states describing the
systems held by Alice and Bob can be taken to be Bell diagonal for the following reason. 
First, Alice and Bob discard all data from mismatched bases, and they 
can assume the worst case scenario which is that the corresponding outcomes are completely uncorrelated. 
This implies 
$\tr[\rho^{AB}(\sigma_i^A\otimes \sigma_j^B)]=0$ for $i\neq j$, $i,j\neq 0$, 
where $\sigma_i$ are the Pauli operators. 
Further, Alice and Bob randomly (but jointly) decide
which state in each basis corresponds to which bit value, so the correlations in each basis are 
characterized by a single error rate $q_j=1-p_0 - p_j$ for $j\in\{1,2,3\}$. 
From this condition, it follows that 
$\tr[\rho^{AB} (\sigma_i^{A} \otimes \mathbbm{1}^B)] = 
\tr[\rho^{AB} (\mathbbm{1}^A \otimes \sigma_i^{B})] =0$ for $i\neq 0$, 
and this leaves only $\sigma_i\otimes \sigma_i$, 
and it is easy to verify that this means the state is Bell diagonal. 
In the six-state scheme, there are three bases, 
so the corresponding error rates determine the Bell-diagonal state completely. 
In the BB84 case, the error rate in the $y$ basis is not known, 
leaving an equivalence class of possible states.

The Bell-diagonal states have a number of appealing and useful properties. 
For instance, it is possible to reduce any bipartite qubit state to the Bell-diagonal form by ``twirling,'' 
choosing a Pauli $\sigma_i$ and applying $\sigma_i^A \otimes \sigma_i^B$ on the state \cite{bennett96b}. 
Generic two-qubit states can also be filtered to the Bell-diagonal form with a two-side filtering \cite{verstraete01a}. 
Finally, any two-qubit state where both reduced states are maximally mixed 
is Bell-diagonal with the right choice of local basis \cite{horodecki96c}, 
and a local change of basis can also rearrange the $p_j$ in any order.

For our purposes, a parametrization different from the $p_j$ will be useful. 
The analysis of both symmetric extension and key distillation is simplified using the following parameters:
\begin{subequations}
\label{eq:alpha-transformation}
\begin{align}
 \alpha_0 &= p_I + p_x + p_y + p_z,\\
 \alpha_1 &= p_I - p_x - p_y + p_z,\\
 \alpha_2 &= \sqrt{2} ( p_I - p_z ),\\
 \alpha_3 &= \sqrt{2} ( p_x - p_y ),
\end{align}
\end{subequations}
which gives the inverse transformation
\begin{subequations}
\begin{align}
 p_I &= \tfrac{1}{4}(\alpha_0 + \alpha_1 + \sqrt{2} \alpha_2),\\
 p_x &= \tfrac{1}{4}(\alpha_0 - \alpha_1 + \sqrt{2} \alpha_3),\\
 p_y &= \tfrac{1}{4}(\alpha_0 - \alpha_1 - \sqrt{2} \alpha_3),\\
 p_z &= \tfrac{1}{4}(\alpha_0 + \alpha_1 - \sqrt{2} \alpha_2).
\end{align}
\end{subequations}
Because of normalization, $\alpha_0 = 1$ for all probability vectors. 
So all Bell-diagonal states are uniquely defined by the coordinates $(\alpha_1,\alpha_2,\alpha_3)$. 
The maximally entangled states are in these coordinates 
$\ket{\Phi^\pm}$: $(1,\pm \sqrt{2}, 0)$ and $\ket{\Psi^\pm}$: $(-1, 0, \pm \sqrt{2})$. 
The convex hull of these four points is a tetrahedron, 
which represents the set of Bell-diagonal states.
This region is defined by the four inequalities
\begin{equation}
  \label{eq:alphaconditions}
  \alpha_1 \pm \sqrt{2} \alpha_2 \geq -1 \quad \text{and} \quad -\alpha_1 \pm \sqrt{2} \alpha_3 \geq -1,
\end{equation}
each corresponding to a particular eigenvalue being non-negative.

\subsection{Formulation as a semidefinite program (SDP)}
\label{ssec:sdpformulation}

Recall that a state $\rho^{AB}$ has a symmetric extension if there exists a state $\rho^{ABB'}$ which is such that 
$\tr_{B'}[\rho^{ABB'}] = \rho^{AB}$ and 
$V_{BB'} \rho^{ABB'} V_{BB'}^\dagger = \rho^{ABB'}$, where $V_{BB'}$ 
is the unitary operation swapping $B$ and $B'$. 
The question of whether or not $\rho^{AB}$ has a symmetric extension can be formulated as a semidefinite program (SDP) \cite{vandenberghe96a,boyd}: 
a convex optimization of a linear function over the convex cone of positive matrices. 
These can be efficiently solved numerically for low-dimensional systems using
interior point algorithms~\cite{doherty02a,doherty04a}. 
The following discussion is adapted from~\cite{doherty04a}. 
Consider the following maximization, a semidefinite program:
\begin{eqnarray}
\label{eq:sdpprimal}
\text{maximize} & 1-\tr[X^{ABB'}],\nonumber\\
\text{subject to} & \tr\,[ \widetilde{L}_i^{ABB'}X^{ABB'}]=\tr[L_i^{AB}\rho^{AB}],\nonumber\\
& X^{ABB'}\geq 0.
\end{eqnarray}
The free variable to be optimized, $X^{ABB'}$, is an 
operator on $\mathcal{H}_A\otimes \mathcal{H}_B\otimes \mathcal{H}_{B'}$, 
and $\{L_i^{AB}\}$ is a basis for traceless operators on $\mathcal{H}_A\otimes \mathcal{H}_B$. 
Further, $\widetilde{L}_i^{ABB'}:=\texttt{Sym}_{BB'}(L_i^{AB}\otimes \mathbbm{1}^{B'})$ for 
$\texttt{Sym}_{BB'}$ the quantum operation symmetrizing systems $B$ and $B'$, 
$\texttt{Sym}_{BB'}(M^{ABB'}) := ( M^{ABB'} + V_{BB'} M^{ABB'} V_{BB'}^\dagger)/2$ using the swap operator  $V_{BB'}$. 

% WRONG!!!
%For any operators $M^{AB}$ and  $N^{ABC}$, we have that
%${\rm Tr_C}[(M^{AB} \otimes \mathbbm{1}^C) N^{ABC}] = M^{AB} \tr_C[N^{ABC}]$.
%It is therefore easy to see that the 
%constraint ensures that the traceless part of either marginal of $X^{ABB'}$ over $B$ or $B'$ equals the 
%traceless part of $\rho^{AB}$. 

If the optimum value of the objective function is nonnegative, 
a suitable multiple $t$ $(\geq 0)$ of $\mathbbm{1}^{ABB'}$ can
be added to $X^{ABB'}$ in order to satisfy the normalization condition 
$\tr[X^{ABB'} + t \mathbbm{1}^{ABB'}]=1$,
and the extension is then given by   
$\rho^{ABB'}=\texttt{Sym}_{BB'}(X^{ABB'} + t \mathbbm{1}^{ABB'})$. 
This symmetrization ensures that $V_{BB'} \rho^{ABB'} V_{BB'}^\dagger = \rho^{ABB'}$.
%and the constraints on $X^{ABB}$ in the SDP ensure that 
%$\tr_{B'}[\rho^{ABB'}] = \rho^{AB}$.
To see that the constraints on $X^{ABB'}$ in the SDP ensure that 
$\tr_{B'}[\rho^{ABB'}] = \rho^{AB}$, we use the facts that
for any operators $M^{AB}$ and  $N^{ABC}$, we have that
$M^{AB} \tr_C[N^{ABC}] = \tr_C[(M^{AB} \otimes \mathbbm{1}^C) N^{ABC}]$ 
and that for any $P^{ABC}$ and $Q^{ABC}$,  
$\tr[ P^{ABC}\texttt{Sym}_{BC}(Q^{ABC}) ] = \tr[ \texttt{Sym}_{BC}(P^{ABC})Q^{ABC} ]$.
We then get that for all $i$,
\begin{align*}
&\tr[L_i^{AB} \tr_{B'}[\texttt{Sym}_{BB'}(X^{ABB'})]]\\
& = \tr[(L_i^{AB} \otimes \mathbbm{1}^{B'}) \texttt{Sym}_{BB'}(X^{ABB'})]\\
& = \tr[\texttt{Sym}_{BB'}(L_i^{AB} \otimes \mathbbm{1}^{B'}) X^{ABB'}]\\
& = \tr[L_i^{AB} \rho^{AB}],
\end{align*}
where the last equality is from the constraint of the SDP \eqref{eq:sdpprimal}.
Since $\{L_i^{AB}\}$ is a basis for the traceless operators on $\mathcal{H}_A\otimes \mathcal{H}_B$, we therefore have that the traceless part of 
$\tr_{B'}[\texttt{Sym}_{BB'}(X^{ABB'})]$ is equal to the traceless part of  $\rho^{AB}$.
Also, $\tr[\rho^{ABB'}] = \tr[\rho^{AB}] = 1$, so $\tr_{B'}[\rho^{ABB'}] = \rho^{AB}$.

If the maximum value is negative, no 
positive semidefinite extension can be constructed  
because if $\rho^{ABB'}$ were a symmetric extension of $\rho^{AB}$, 
the choice $X^{ABB'} = \rho^{ABB'}$ would satisfy the constraints and give the objective function a value of 0.

To every SDP, there is an associated dual SDP 
which for the symmetric extension problem is somewhat easier to work with. 
The dual of Eq.~\eqref{eq:sdpprimal} is
\begin{eqnarray}
\label{eq:sdpdual}
\text{minimize} &\tr[K^{AB}\rho^{AB}],\nonumber\\
\label{eqn:primalproblem}
\text{subject to} & \widetilde{K}^{ABB'}\geq 0,
\end{eqnarray}
where $K^{AB}:=\mathbbm{1}^{AB}+\sum_{j}l_j L_j^{AB}$, 
$\widetilde{K}^{ABB'} := \texttt{Sym}_{BB'}(K^{AB}\otimes \mathbbm{1}^{B'})$, 
and $l_j$ are free variables to be optimized. 
We refer to this optimization as the \emph{dual problem} 
and the original optimization \eqref{eq:sdpprimal} as the \emph{primal problem}. 
Should $\rho^{AB}$ have a symmetric extension, 
$\tr[K^{AB}\rho^{AB}]=\tr[\widetilde{K}^{ABB'}\rho^{ABB'}]\geq 0$,
since the trace of the product of two positive operators is non-negative.
Thus a sufficient condition for $\rho^{AB}$ \emph{not} 
to be extendible is for
the minimum of  $\tr[K^{AB}\rho^{AB}]$ to be negative. 
As we discuss below, this condition is also necessary,
due to a property known as \emph{strong duality}.

\emph{Weak duality} holds that the optimum value of the primal problem is always less than the optimum of the dual, 
which follows from the positivity constraints,
\begin{multline}
\label{eqn:weakduality}
\tr[K^{AB}\rho^{AB}]-(1-\tr[X^{ABB'}])\\
=\tr[\widetilde{K}^{ABB'}X^{ABB'}]
\geq 0, 
\end{multline}
where the equality follows from 
$\tr[K^{AB}\rho^{AB}] 
= \tr[\widetilde{K}^{ABB'}\rho^{ABB'}]
= \tr[\widetilde{K}^{ABB'}(X^{ABB'} + t \mathbbm{1}^{ABB'})]$
and
$1-\tr[X^{ABB'}] = \tr[t \mathbbm{1}^{ABB'}]$.

Strong duality is the statement that the optimum values of the primal
and dual problems are equal. Sufficient conditions for strong duality
are known.
In particular, a semidefinite program is said to be \emph{strictly
  feasible} if the
constraints can be satisfied by a matrix that is positive definite rather than
just positive semidefinite. The strict feasibility of either the
primal or dual semidefinite programs is sufficient to guarantee strong
duality (theorem 3.1~\cite{vandenberghe96a}). If both the primal and
the dual are strictly feasible then we
are also guaranteed that there exist matrices $X^{ABB'}_\textrm{opt}$ and $\widetilde{K}^{ABB'}_\textrm{opt}$ that satisfy
the constraints and attain the optimum of the primal and dual program,
respectively, (theorem 3.1~\cite{vandenberghe96a}). 

The dual problem is obviously strictly feasible just by taking  $K^{AB}=\mathbbm{1}^{AB}$. 
On the primal side, note that there must be some, not necessarily
positive, $X^{ABB'}$ 
meeting the constraints, since these underdetermine the components. 
As the constraints only involve the traceless part of $X^{ABB'}$,
a suitable 
multiple of the identity $\mathbbm{1}^{ABB'}$ can always be added to
ensure positivity. 
From strict feasibility, we obtain the sufficiency condition that $\min(\tr[K^{AB}\rho^{AB}])>0$ 
implies $\rho^{AB}$ is extendible\footnote{In this manner, the dual SDP constructs a \emph{witness} for the (lack of) symmetric extension, a Hermitian 
operator $K^{AB}$ which defines a 
hyperplane in the set of positive operators separating the given state $\rho^{AB}$ from the convex 
set of extendible states \cite{horodecki96a}. 
Strict feasibility implies that we only need to consider 
Hermitian $K^{AB}$ for which $\widetilde{K}^{ABB'}\geq 0$.}.
Moreover, when the optima are equal, $\tr[\widetilde{K}^{ABB'}_\textrm{opt}X^{ABB'}_\textrm{opt}]=0$, 
and hence $\widetilde{K}^{ABB'}_\textrm{opt}X^{ABB'}_\textrm{opt}=0$. 
This condition is termed
\emph{complementary slackness} and will play an important role in the analytical solution.

\subsection{Simplifying the SDP for Bell-diagonal states}

Now consider the dual form of the SDP, [Eq.~(\ref{eq:sdpdual})]. 
By exploiting the symmetry of the problem, we can find an analytic solution. 
The method for dealing with symmetry follows the general
prescription of 
Gatermann and Parrilo~\cite{gatermann}, but takes advantage of several special
properties of this problem. Because $\rho^{AB}$ is Bell diagonal,
it is invariant under conjugation by Pauli operators $\sigma_i \otimes \sigma_i$.
This induces a symmetry of the objective function, since 
$\tr[(\sigma_i \otimes \sigma_i)K(\sigma_i \otimes \sigma_i)^\dagger\rho]=\tr[K\rho]$ for any 
$i \in \{ 0, 1, 2, 3 \}$.
Moreover, the constraint 
$\widetilde{K}^{ABB} = \texttt{Sym}_{BB'}(K^{AB}\otimes \mathbbm{1}^{B'}) \geq 0$ 
is equivalent to 
$\texttt{Sym}_{BB'}((\sigma_i \otimes \sigma_i)K^{AB}(\sigma_i \otimes \sigma_i)^\dagger \otimes \mathbbm{1}^{B'}) \geq 0$.
Hence, the set of allowable $K$ is invariant under 
arbitrary conjugation by Pauli operators and since they all yield the same 
value of the objective function, we can focus on those formed by the convex combination  
$\bar{K}=\frac{1}{4}\sum_{i=0}^3 (\sigma_i \otimes \sigma_i)K(\sigma_i \otimes \sigma_i)^\dagger$
without loss of generality. 
Since $\bar{K}$ is a ``twirl'' of $K$, it is also Bell diagonal: $\bar{K}=\sum_j k_j \ketbra{\beta_j}$. 
The $k_j$ satisfy 
$\sum_j k_j =1$, since $\tr[\bar{K}] = \tr[(\sigma_i \otimes \sigma_i)K(\sigma_i \otimes \sigma_i)^\dagger] = 1$, 
but not necessarily $k_j \geq 0$.
%Writing $\bar{K}=\sum_{j}k_{j}\ketbra{\beta_j}$, we have that $\rho^{AB}$ has a symmetric extension whenever $\min\sum_{j}k_{j}p_{j}\geq 0$ subject to $\texttt{Sym}_{BB'}(\bar{K}\otimes \mathbbm{1})\geq 0$.
This simplifies the objective function $\tr[K^{AB}\rho^{AB}]$ to $\sum_{j}k_{j}p_{j}$ with the additional constraint $\sum_j k_j = 1$.

Next, we would like to use the symmetry of $\bar{K}$ to simplify the
constraint $\widetilde{K}^{ABB'} \geq 0$. 
For readability, we will---in
this and the next two paragraphs---write the Pauli operators as 
$X$, $Y$, and $Z$, 
and tensor products such as $\mathbbm{1} \otimes \sigma_x \otimes
\sigma_z$ as $IXZ$. 
Observe that $\widetilde{K}$ 
inherits invariance under the operators $XXX$ and $ZZZ$ from
$\bar{K}$.  We can simplify the calculation by observing that
$XXX$ and $ZZZ$ are logical operators for the bit-flip code. Because
of the symmetry, it will be necessary for $\widetilde{K}$ to be
proportional to the
identity on the code space. $\widetilde{K}$ has a symmetry under
swapping $B$ and $B'$ that we will also wish to take advantage of. 
We can proceed by identifying three ``logical'' or encoded qubits $F$,
$G$, and $H$ on the Hilbert space 
$\mathcal{H}_A\otimes\mathcal{H}_B\otimes\mathcal{H}_{B'}$, 
such that the form of $\widetilde{K}$ is simpler when expressed 
in the computational basis of $\mathcal{H}_F\otimes\mathcal{H}_G\otimes\mathcal{H}_{H}$.
The encoded $X$- and $Z$-operators on the logical qubits are 
%$(X_F,Z_F) := (XXX,ZZZ)$, $(X_G,Z_G) := (XIX,ZZI)$, and $(X_H,Z_H) := (XXI,ZIZ)$. 
\begin{align*}
 X_F &:= XXX, & X_G &:= XIX, & X_H &:= XXI \\
 Z_F &:= ZZZ, & Z_G &:= ZZI, & Z_H &:= ZIZ.
\end{align*}
Note that with these definitions swapping $B$ and $B'$ induces a swap
on $G$ and $H$.
It is simple to verify that the Pauli operators on different logical qubits commute and that $X$ and $Z$ anticommute on the same logical qubit. 
They therefore define a valid encoding, and the encoded product vectors $\ket{ijk}_{FGH}$ are the eight simultaneous eigenvectors of the encoded $Z$ operators,
% Product vectors in ABC are eigenvectors of Z_F, Z_G and Z_H
% U_{ABC->FGH} defined by permutations (001,101,010,110)(100,111)(000)(011)
\begin{equation}
\label{eq:ABBFGH}
\begin{aligned}
  \ket{000}_{FGH} & = \ket{000}_{ABB'}, & \ket{100}_{FGH} & = \ket{111}_{ABB'},\\
  \ket{001}_{FGH} & = \ket{110}_{ABB'}, & \ket{101}_{FGH} & = \ket{001}_{ABB'},\\
  \ket{010}_{FGH} & = \ket{101}_{ABB'}, & \ket{110}_{FGH} & = \ket{010}_{ABB'},\\
  \ket{011}_{FGH} & = \ket{011}_{ABB'}, & \ket{111}_{FGH} & = \ket{100}_{ABB'}.
\end{aligned}
\end{equation}

Since $\widetilde{K}$ 
is invariant under the operators $XXX$ and $ZZZ$, we can immediately
infer that $\widetilde{K}\simeq\mathbbm{1}_F\otimes
\widetilde{K}'_{GH}$. 
Furthermore, $\widetilde{K}$ is by definition invariant under swapping the $BB'$ systems, 
and swapping $BB'$ is the same as swapping $GH$.
This means that $\widetilde{K}'_{GH}$ must be block diagonal with
the support on the triplet and singlet subspaces.
Since $\bar{K}$ is Bell diagonal and 
%$\widetilde{K}$ is linear in $\bar{K}$, 
$\texttt{Sym}_{BB'}$ is a linear superoperator,
we can write $\widetilde{K} = \sum_j k_j \texttt{Sym}_{BB'}(\ketbra{\beta_{j}}\otimes\mathbbm{1}_{B'})$.
Converting the terms $\texttt{Sym}_{BB'}(\ketbra{\beta_{j}}\otimes\mathbbm{1}_{B'})$ 
into operators on the logical qubits can be accomplished by writing it out in the computational basis and using the relations \eqref{eq:ABBFGH}.

Alternatively, the conversion can be done by noticing that $ZZI$ and $XXI$ are encoded $Z$ and $X$ operators for logical qubits $G$ and $H$, respectively. 
Thus, 
$\proj{\Phi^+}_{AB}\otimes\mathbbm{1}_{B'}$ is on the +1 eigenspace
of both $Z_G$ and $X_H$, i.e., 
$\proj{\Phi^+}_{AB}\otimes\mathbbm{1}_{B'} \simeq \mathbbm{1}_F\otimes\ketbra{0{+}}_{GH}$. 
For $\ket{\Phi^-}_{AB}$, we use that 
$\proj{\Phi^-}_{AB}\otimes\mathbbm{1}_{B'}$ = $(ZII)(\proj{\Phi^+}_{AB}\otimes\mathbbm{1}_{B'})(ZII)^\dagger$, 
and since $ZII$ commutes (anticommutes) with $ZZI$ ($XXI$), 
$\proj{\Phi^-}_{AB}\otimes\mathbbm{1}_{B'}$ is on the $+1$ ($-1$) eigenspace of $ZZI$ ($XXI$). 
Therefore, $\proj{\Phi^-}_{AB}\otimes\mathbbm{1}_{B'} \simeq \mathbbm{1}_F\otimes\ketbra{0{-}}_{GH}$. 
Similarly, $\ket{\Psi^+}_{AB}$ and $\ket{\Psi^-}_{AB}$ correspond to $\ket{1+}_{GH}$ and $\ket{1-}_{GH}$, respectively.
Applying the swap symmetrization $\texttt{Sym}_{BB'}$ to $\ketbra{\beta_{j}}\otimes\mathbbm{1}_{B'}$ 
is simple given this concrete representation, and the results have the following
form. 
First, each of the terms has the form 
$\texttt{Sym}_{BB'}(\ketbra{\beta_{j}}\otimes\mathbbm{1}_{B'})\simeq \frac{1}{8}\mathbbm{1}_F\otimes \left(R_j\oplus \Psi^{-}_{GH}\right)$, 
where $R_j$ has support only on the triplet subspace. 
A simple calculation shows that the $R$ matrices are given by
\begin{align}
\label{eq:R1}
        R_{\Phi^{\pm}}&=\begin{pmatrix}
                2 & \pm \sqrt{2} & 0\\
\pm \sqrt{2} & 1 & 0\\
0 & 0 & 0
        \end{pmatrix},\\
\label{eq:R2}
R_{\Psi^{\pm}}&=\begin{pmatrix}
        0 & 0 & 0\\
0 & 1 & \pm \sqrt{2}\\
0 & \pm \sqrt{2} & 2
\end{pmatrix},
\end{align}
in the basis $\{\ket{00},\ket{\Psi^+},\ket{11}\}$. 

The semidefinite program \eqref{eqn:primalproblem} now becomes
\begin{equation}
  \label{eq:SDP3}
\begin{aligned}
  \textrm{minimize} \quad & {\textstyle \sum_{i=0}^3 k_i p_i}, \\
\textrm{subject to} \quad & {\textstyle \sum_{i=0}^3 k_i R_i \geq 0}, \\
& {\textstyle \sum_{i=0}^3 k_i =1},
\end{aligned}
\end{equation}
where $\{p_i\}$ ($\{k_i\}$) are the eigenvalues of $\rho$ ($\bar{K}$).  
The latter constraint can be eliminated by a further change of variables according to Eq.~\eqref{eq:alpha-transformation} for both $p_i$ and $k_i$, so that $p \rightarrow \alpha$ and $k \rightarrow x$. 
The latter constraint now becomes simply $x_0 = 1$, and only $x_1$, $x_2$, and $x_3$ remain as free variables.
The objective function becomes 
%$\sum_{i=0}^3 k_i p_i = $
$(x_0 \alpha_0 + \sum_{j=1}^3 x_j \alpha_j)/4$.
Instead of minimizing this directly, we multiply by 4 and subtract the constant term $x_0 \alpha_0 =1$. 
This gives us the following much-simplified dual SDP which is equivalent to SDP \eqref{eq:SDP3} except for a rescaled and shifted objective function,
\begin{equation}
\label{eq:simpdual}
\begin{aligned}
  \textrm{minimize} \quad & \sum_{j=1}^3x_j\alpha_j, \\
\textrm{subject to} \quad & F(x)=F_0+\sum_{i=1}^3 x_i F_i \geq 0,
\end{aligned}
\end{equation}
using the matrices
\begin{eqnarray*}
         F_0  &=&  \left(
\begin{array}{ccc}
  1 & 0 & 0  \\
  0 & 1 & 0  \\
  0 & 0 & 1 
\end{array}
\right),\quad F_1 = \left(
\begin{array}{ccc}
  1 & 0 & 0  \\
  0 & 0 & 0  \\
  0 & 0 & -1 
\end{array}
\right),\\
F_2 &=& \left(
\begin{array}{ccc}
  0 & 1 & 0  \\
  1 & 0 & 0  \\
  0 & 0 & 0 
\end{array}
\right),\quad
F_3 =  \left(
\begin{array}{ccc}
  0 & 0 & 0  \\
  0 & 0 & 1  \\
  0 & 1 & 0 
\end{array}
\right).
\end{eqnarray*} 
If the minimum value of the objective function is greater than or equal to $-1$, the state has a symmetric extension. Because of the minimization, finding an $x$ that satisfies the constraints and such that the objective function is less than $-1$ is sufficient to show that the state does not have a symmetric extension.

We can find the simplified form of the primal problem by taking the
dual of the SDP~\eqref{eq:simpdual} 
\begin{equation}
\label{eq:simpprimal}
\begin{aligned}
\textrm{minimize} \quad & \tr[Z],\\
\textrm{subject to}\quad & \tr[F_iZ]={\alpha_i},\\
& Z\geq 0,
\end{aligned}
\end{equation}
where again the state has a symmetric extension when Tr$[Z^*]\leq 1$. 
We use $^*$ throughout to denote an optimal value of a variable. 
Finding any $Z$ that satisfies the constraints and has trace less than or equal to $1$ is sufficient to show that the state has a symmetric extension.

\subsection{Analytical solution of the SDP}

In this section we will solve the simplified semidefinite program using both the primal form \eqref{eq:simpprimal} and the dual form \eqref{eq:simpdual}.
For the states which have a symmetric extension, 
we prove this by finding an explicit $Z$ with 
$\tr[Z] \leq 1$ 
which satisfies the constraints of the SDP \eqref{eq:simpprimal}.
When the state has no symmetric extension, 
this can be proven by finding an $x$ such that
the constraints of \eqref{eq:simpdual} is satisfied 
and $\sum_{j=1}^3 x_j \alpha_j \leq -1$, 
but we will not use this. 

As shown in Sec.~\ref{ssec:sdpformulation}, 
the optima $Z^*$  and $F(x^*)$ from the primal and dual problems
obey the complementary slackness condition 
\begin{equation}
  \label{eq:slackness}
  F(x^*)Z^*=0,
\end{equation}
and it is this condition that 
allows us to solve the semidefinite program analytically and prove that certain states do not have a symmetric extension.
The first simplification we get from condition \eqref{eq:slackness} is that 
$\rank[F(x^*)] + \rank(Z^*) \leq 3$
since $F(x^*)$ and $Z^*$ must have support on orthogonal subspaces.
Since both $F(x^*) = 0$ and $Z^* = 0$ are excluded by the constraints, 
at least one of $F(x^*)$ and $Z^*$ must have rank one.

The solution will proceed as follows.
We first consider $Z$ of rank one.
This will give us a sufficient condition for a symmetric extension.
We then consider the case when this condition is not satisfied.
Under the assumption that the state still has a symmetric extension, we use complementary slackness to show that there can only be four possible $Z^*$. 
If none of these candidates satisfy $Z \geq 0$, we get a contradiction and the state 
cannot have a symmetric extension.
It turns out that the candidates all satisfy $\tr[Z] \leq 1$, though, 
so if one of them also is positive semidefinite, it also proves that the state has a symmetric extension.

Start by finding the possible values for the objective function when $Z$ is rank one.
From the constraints $\tr[F_iZ]={\alpha_i}$ of the primal problem, $Z$ has the form 
\begin{equation}
\label{eq:Zform}
  Z =\frac{1}{2} 
\begin{pmatrix}
  2(\alpha_1+z_{33}) & \alpha_2 & 2z_{13}  \\
   \alpha_2 & 2z_{22} &  \alpha_3  \\
  2z_{13} & \alpha_3 & 2z_{33}
\end{pmatrix}.
\end{equation}
The objective function is the trace of this matrix, so we want to determine $z_{22}$ and $z_{33}$ from the rank-one condition.
Since $Z$ is real and symmetric, we can parametrize its eigenvector
with three real numbers $a_i$. This gives
an alternative characterization of $Z$, 
\begin{equation}
\label{eq:rankone}
  Z = \left(
\begin{array}{ccc}
  a_1^2 & a_1 a_2 & a_1 a_3  \\
  a_1 a_2 & a_2^2 & a_2 a_3  \\
  a_1 a_3 & a_2 a_3 & a_3^2
\end{array}
\right),
\end{equation}
and we can solve the problem by equating these.  
Taking the ratio of the 1,2 and 2,3 elements, we get
$a_1/a_3 = \alpha_2 /
\alpha_3$ when $a_2, a_3, \alpha_3\neq 0$. 
The ratio of the 1,1 and 3,3 elements is the square of this, 
which implies that $z_{33}=a_3^2=\alpha_1\alpha_3^2/(\alpha_2^2-\alpha_3^2)$.
Now use the fact that the square of the 2,3 element equals $z_{22}z_{33}$ to find $z_{22}=(\alpha_2^2-\alpha_3^2)/4\alpha_{1}$. 
The objective function is then 
\begin{equation}
\label{eq:objfunZrankone}
        \tr[Z]=\frac{ (\alpha_2^2 - \alpha_3^2)^2 + 4 \alpha_1^2 (\alpha_2^2 + \alpha_3^2)}{4 \alpha_1 (\alpha_2^2-\alpha_3^2)},
\end{equation}
and since it is fixed by the state (and the requirement that $Z$ be rank one), no minimization is required.
If this expression is less than or equal to 1, the state has a symmetric extension since $\tr[Z^*] \leq \tr[Z]$. 
If the value is greater than 1, we cannot conclude yet since it could be that $Z^*$ is of rank two and has trace less than or equal to one. 
Thus,
\begin{equation}
\label{eq:rankoneZsufcond}
 4 \alpha_1 (\alpha_2^2-\alpha_3^2) - (\alpha_2^2 - \alpha_3^2)^2 - 4 \alpha_1^2 (\alpha_2^2 + \alpha_3^2) \geq 0
\end{equation}
is a sufficient but not necessary condition for the state to have a symmetric extension.

If Eq.~\eqref{eq:rankoneZsufcond} is satisfied, we know that the state has a symmetric extension, so for the rest of this section we assume that it is not.
For a contradiction (in some cases), we now assume that the state has a symmetric extension.
This means that $\rank(Z^*) = 2$ and because of complementary slackness 
$\rank[F(x^*)] = 1$.
We therefore want to find out for what possible $x$ we get a rank one $F(x)$. 
The dual problem \eqref{eq:simpdual} gives us the form of $F(x)$,
\begin{equation*}
        F(x)=\begin{pmatrix}
                1+ x_1 & x_2 & 0\\
x_2& 1 & x_3\\
0 & x_3 & 1-x_1         \end{pmatrix}.
\end{equation*} 
In this case we proceed as before, expressing $F(x)$ also as a projection operator of the form $\eqref{eq:rankone}$ and using relations between the matrix elements.
From the 1,3 element, it is clear that either $a_3$ or $a_1$ must be zero. 
This zeroes out the first or third column and row, and we immediately obtain $x_1=\pm 1$ and $x_3=0$ or $x_2=0$ for the former and latter cases, respectively. 
This leaves a matrix with a non-zero $2 \times 2$ block, which must have determinant zero. 
From this we get $x_2^2 = 2$ and $x_3^2 = 2$ in the two cases, so $x = (x_1,x_2,x_3)$ can only take one of the four values $(1,\pm \sqrt{2}, 0), (-1,0, \pm \sqrt{2})$.
%This corresponds to the four possibilities $k_j = \delta_{jm}$ in \eqref{eq:SDP3}, and the
%corresponding matrices $F(x)$ are exactly the $R$ matrices from \eqref{eq:R1} and \eqref{eq:R2}.
The corresponding four values of the objective function in SDP \eqref{eq:simpdual} are
\begin{equation}
 \alpha_1 \pm \sqrt{2} \alpha_2 \quad \text{and} \quad -\alpha_1 \pm \sqrt{2} \alpha_3.
\end{equation}
If any of these would be less than $-1$, we would be able to exclude the possibility of a symmetric extension at this point. 
However, this is not possible for any states, 
since the four inequalities \eqref{eq:alphaconditions} defining the border of the set of Bell-diagonal states are saying exactly that 
these four values are greater than or equal to $-1$.

The four possible candidates for $x^*$ cannot by themselves contradict our assumption of a symmetric extension for possible values of $\alpha_i$.
However, under this assumption one of these candidates must be optimal. 
There must, therefore, be a complementary optimal $Z^*$ of the primal problem for which
the complementary slackness condition \eqref{eq:slackness} is satisfied.
For each of the four possible $x^*$, we can impose the complementary slackness condition $F(x) Z = 0$ to a $Z$ of the form \eqref{eq:Zform}, 
and check if the resulting $Z$ can be positive semidefinite as required by the SDP conditions. 
For the two vectors $x = (1,\pm \sqrt{2}, 0)$, this gives the two possible matrices
\begin{equation}
\label{eq:rank2Z}
        Z=\frac{1}{2\sqrt{2}}\begin{pmatrix}
                \mp\alpha_2 & \sqrt{2}\alpha_2& \mp\alpha_3\\
\sqrt{2}\alpha_2 & \mp 2\alpha_2 & \sqrt{2}\alpha_3\\
\mp\alpha_3 & \sqrt{2}\alpha_3 & -2\sqrt{2}\alpha_1\mp \alpha_2
        \end{pmatrix}.
\end{equation}
Since the second column is proportional to the first, 
the matrix is positive semidefinite if and only if the lower right $2 \times 2$ block is. 
This is positive semidefinite if and only if both the determinant and one of the diagonal elements are non-negative. 
The determinant is in this case proportional to $2(\alpha_2^2-\alpha_3^2)\pm4\sqrt{2}\alpha_1\alpha_2$, 
so the matrix is positive semidefinite if and only if 
$\mp \alpha_2 \geq 0$ 
and 
$\alpha_2^2-\alpha_3^2 \pm 2\sqrt{2}\alpha_1\alpha_2 \geq 0$.
The possible matrices for $x = (-1,0, \pm \sqrt{2})$ 
are the matrices we get from Eq.~\eqref{eq:rank2Z} by interchanging the first and third rows and columns and making the substitutions 
$\alpha_2 \leftrightarrow \alpha_3$, $\alpha_1 \leftrightarrow - \alpha_1$. 
The positivity conditions are 
$\mp \alpha_3 \geq 0$ and $\alpha_3^2-\alpha_2^2 \mp 2\sqrt{2}\alpha_1\alpha_3 \geq 0$. 
Thus, if the state does not satisfy condition \eqref{eq:rankoneZsufcond}, 
and also none of the four positivity constraints,
\begin{subequations}
\label{eq:rank2Zposconstraints}
\begin{align}
\label{eq:rank2Zposconstraint1}
 \alpha_2^2-\alpha_3^2  \pm 2\sqrt{2}\alpha_1 \alpha_2 \geq 0 \quad & \text{and} \quad  \mp \alpha_2 \geq 0, \\
\label{eq:rank2Zposconstraint2}
 \alpha_3^2-\alpha_2^2  \mp 2\sqrt{2}\alpha_1 \alpha_3 \geq 0 \quad & \text{and} \quad  \mp \alpha_3 \geq 0
\end{align}
\end{subequations}
we cannot have $\rank[F(x^*) = 1]$, so our assumption that the state has a symmetric extension is contradicted.

If on the other hand, one or more of the constraints are satisfied,
there is no contradiction and the state could have a symmetric extension. 
Actually, we can use the $Z$ which satisfies $Z \geq 0$ to prove that a symmetric extension exists.
Taking the trace in Eq.~\eqref{eq:rank2Z} gives 
$\tr[Z] = -\alpha_1 \mp \sqrt{2} \alpha_2 \leq 1$, 
where the inequality follows from the first two border inequalities
in Eq.~\eqref{eq:alphaconditions}.
For the $Z$ corresponding to $x = (-1,0, \pm \sqrt{2})$, we can show $\tr[Z] \leq 1$ by using the other two border inequalities. 
The $Z$ which is positive semidefinite will therefore satisfy all the constraints of the primal SDP \eqref{eq:simpprimal}, and since it gives a value of the objective function which is less than or equal to $1$, the state must have a symmetric extension.

\begin{figure*}
%  \texttt{Picture commented out until graphics files are added to SVN}
%  \resizebox{\textwidth}{!}{\includegraphics{gfx/newcompound}}
  \includegraphics{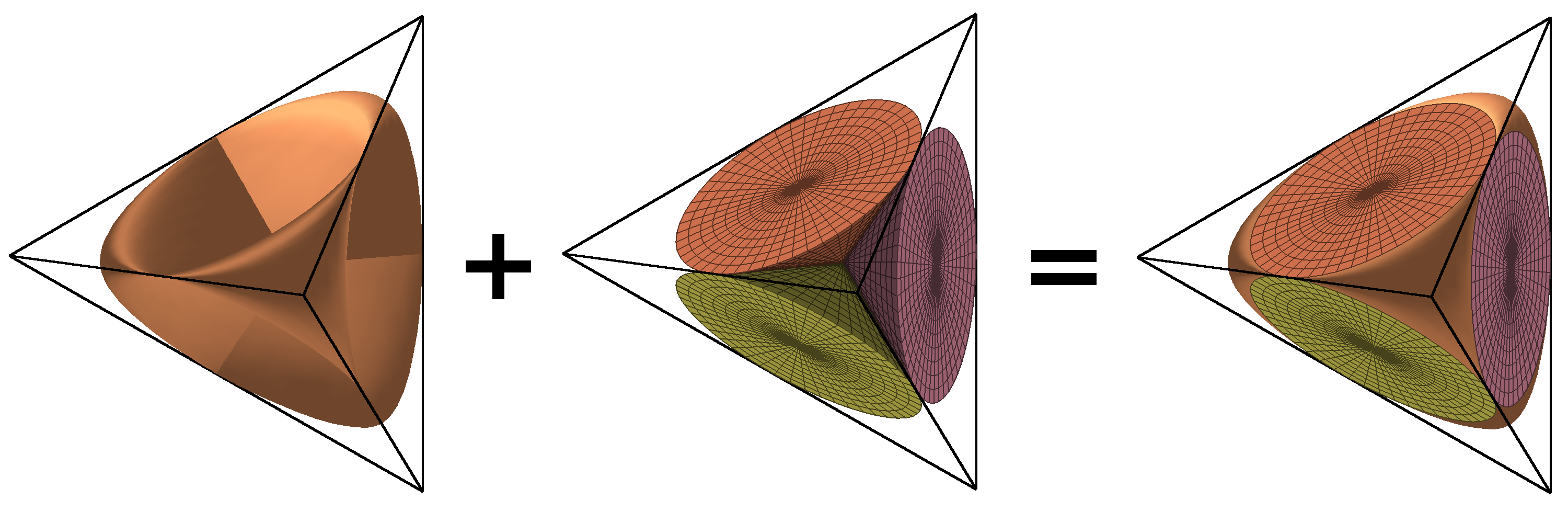}
\caption{\label{fig:extstates}The set of Bell-diagonal states that satisfies the rank-one $Z$ condition \eqref{eq:rankoneZsufcondsum} (left), 
rank-two $Z$ conditions \eqref{eq:rank2Zposconstraint1sum} or \eqref{eq:rank2Zposconstraint2sum} (center), and the union of the two (right). 
The figures have a maximally entangled state on each vertex and the surfaces have the symmetry of the tetrahedron.}
\end{figure*}

Altogether, we have shown that if any of the conditions 
\eqref{eq:rankoneZsufcond}, \eqref{eq:rank2Zposconstraint1}, or \eqref{eq:rank2Zposconstraint2} 
are satisfied, the state has a symmetric extension, otherwise it does not. 
Since at least one of $\mp \alpha_2 \geq 0$ always holds, 
we can combine the two options in Eq.~\eqref{eq:rank2Zposconstraint1} into one at the cost of adding an absolute value. 
We can do the same for $\mp \alpha_3 \geq 0$ in Eq.~\eqref{eq:rank2Zposconstraint2}
and combining everything we get that a state has symmetric extension if and only if one or more of the following three inequalities hold:
\begin{subequations}
\label{eq:conditionsummary}
\begin{gather}
\label{eq:rankoneZsufcondsum}
 4 \alpha_1 (\alpha_2^2-\alpha_3^2) - (\alpha_2^2 - \alpha_3^2)^2 - 4 \alpha_1^2 (\alpha_2^2 + \alpha_3^2) \geq 0, \\
\label{eq:rank2Zposconstraint1sum}
 \alpha_2^2-\alpha_3^2 - 2\sqrt{2}\alpha_1|\alpha_2| \geq 0, \\
\label{eq:rank2Zposconstraint2sum}
 \alpha_3^2-\alpha_2^2 + 2\sqrt{2}\alpha_1|\alpha_3| \geq 0.
\end{gather}
\end{subequations}
The set of Bell-diagonal states with symmetric extension is pictured in Fig.~\ref{fig:extstates}.
Condition \eqref{eq:rankoneZsufcondsum} describes a body that includes the 
symmetric extendible states closest to the maximally entangled states. 
It is, however, not convex. 
The conditions \eqref{eq:rank2Zposconstraint1sum} and \eqref{eq:rank2Zposconstraint2sum} 
describe four cones with vertex at the maximally mixed state 
and a maximal circular base on each face of the tetrahedron. 
The cones fill in the convex hull of the first body, 
so that the body of symmetric extendible states is just 
the convex hull of the body from condition \eqref{eq:rankoneZsufcondsum}.

\section{Thresholds for QKD schemes}
\label{sec:qkd}

\subsection{The Chau protocol}

The two-way procedure to distill secret key from quantum correlations in a prepare and measure scheme invented by Gottesman and Lo \cite{gottesman03a} 
was proven by Chau \cite{chau02a} to work for Bell-diagonal states with error rates that satisfy
\begin{equation} 
\label{eq:chau-condition}
(p_I - p_z)^2 > (p_I + p_z)(p_x + p_y). 
\end{equation}
This corresponds to a quantum bit error rate (QBER) of 27.64\% for the six-state scheme ($p_x = p_y = p_z$) and 20\% for BB84 ($p_x = p_z$, $p_y = 0$). 
In this section, we show that when this condition is satisfied, the procedure breaks the symmetric extension in a finite number of rounds, 
as implied by Chau's result. 
When it is not satisfied, however, the procedure can only output states with symmetric extension, and therefore no key can be distilled. 
This is similar to the analysis by Ac\'in \textit{et al.}~\cite{acin06a}, 
but since we know when a state has a symmetric extension, 
we do not need to construct an explicit attack. 

% Introduce B- and P-steps. Argue that only B-steps can break extension
The procedure works by first applying a number of so-called B-steps (for bit error detection) 
then P-steps (for phase error correction) and in the end a one-way quantum error correcting code. 
The B-step works on two bit pairs. 
On each side, the parity of the bits is computed and compared to the other side. 
If the parity differs, there must have been an error and both pairs are discarded. 
If the parity is equal, the first pair is kept. 
This step requires two-way communication since both parties need to know if they should keep the first pair.
The P-step works on three bit pairs. 
The output bit on each side is the parity of the three bits. 
This does not require any communication at all, 
but it simulates a phase error correction step where two qubits are measured to give a phase error syndrome which is sent from Alice to Bob for comparison.
Alternatively, we can look at it as keeping the two extra qubits on each side in a
\emph{shield} system which limits Eve's knowledge about the \emph{key} system \cite{horodecki05a}.
Irrespective of how we look at it, a P-step does not require communication from Bob to Alice and can therefore not break a symmetric extension.
If states with symmetric extension are to be distilled into secret key, the B-steps must break the symmetric extension, and we will therefore concentrate on these in the following.

% Give recursion formulas for B-steps. 
After a successful round of B-steps, the new error probabilities are \cite{chau02a}:
\begin{subequations}
\label{eq:chau-recursion}
\begin{align}
p_I^\text{out} &= \frac{p_I^2 + p_z^2}{(p_I + p_z)^2+(p_x + p_y)^2},\\
p_x^\text{out} &= \frac{p_x^2 + p_y^2}{(p_I + p_z)^2+(p_x + p_y)^2},\\
\label{eq:chau-recursion-y}
p_y^\text{out} &= \frac{2 p_x p_y}{(p_I + p_z)^2+(p_x + p_y)^2},\\
p_z^\text{out} &= \frac{2 p_I p_z}{(p_I + p_z)^2+(p_x + p_y)^2}.
\end{align}
\end{subequations}
% Define D_C and show that it doubles for every successful step
To quantify how the procedure improves or deteriorates the ability of a state to produce a key, 
as defined by Eq.~\eqref{eq:chau-condition}, 
we define the quantity\footnote{We do not care about the rate, so we expect no relation between $D_C$ and the key rate. It is possible to have arbitrarily high $D_C$ and at the same time arbitrarily low key rate.}
\begin{equation}
\label{eq:chau-distillability}
 D_C := \log_2 \left(  \frac{(p_I - p_z)^2}{(p_I + p_z)(p_x + p_y)} \right).
\end{equation}
% I call this Chau-distillability in lack of a better name, should we use that name?
This quantity is positive on all distillable states, negative on states where $(p_I - p_z)^2 < (p_I + p_z)(p_x + p_y)$, and zero on the border. 
By inserting the recursion relation \eqref{eq:chau-recursion} into Eq.~\eqref{eq:chau-distillability}, 
we see that $D_C$ doubles for every successful B-step, $D_C^\text{out} = 2 D_C$. 
Thus, if the state starts out with negative $D_C$, it will remain negative, 
if it starts out being zero it will remain so, 
and if it starts out being positive it can reach an arbitrary positive value in a finite number of steps. 
We will next show that this allows the procedure to break the symmetric extension when $D_C > 0$ and not otherwise. 
More precisely, we will show that reaching $D_C \geq 2$ is sufficient for breaking the symmetric extension, 
whereas all states with $D_C \leq 0$ have a symmetric extension.

\begin{figure}
\includegraphics{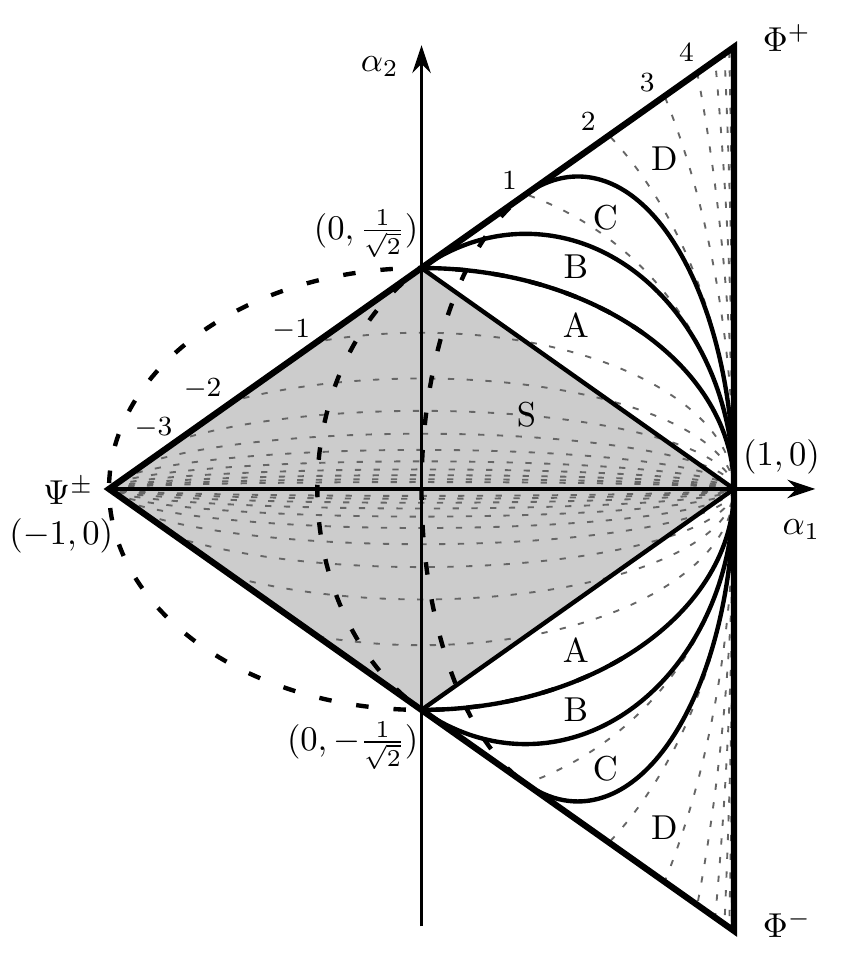} 
\caption{\label{fig:statespace} Plot of the Bell-diagonal state space as a function of $\alpha_1$ and $\alpha_2$, with $\alpha_3$ projected out.
The thin dashed lines indicate the value of $D_C$.
The shaded region S corresponds to separable states (for at least some $\alpha_3$). 
Region A is the set of entangled states for which the B-steps fail to break a symmetric extension. 
The border between A and B corresponds to $D_C = 0$.
Regions A and B together are the entangled states with symmetric extension for all possible values of $\alpha_3$, 
while in region C all the states with $\alpha_3 = 0$ have a symmetric extension; 
but some states with other $\alpha_3$ do not.
In region D, no states have symmetric extension.
The borders between regions B and C and regions C and D both have the shape of ellipses.
The former is described by Eq. \eqref{eq:surfaceellipse}, 
while the latter is described by Eq. \eqref{eq:sixstateellipse}.}
\end{figure}

To show this, we describe the states by the same parameters $\alpha$
that we used in the symmetric extension calculation 
and defined in Eqs.~\eqref{eq:alpha-transformation}.
In these coordinates, 
 $D_C = \log_2 [  2 \alpha_2^2/(1-\alpha_1^2) ]$
does not depend on $\alpha_3$ at all. 
The equations for the surfaces of constant $D_C$ are then
\begin{equation}
\label{eq:chauconst}
 \alpha_1^2 + 2\cdot 2^{-D_C} \alpha_2^2 = 1.
\end{equation}
These are the equations for ellipses with center in the origin, 
constant $\alpha_1$-semiaxis $1$, and $D_C$-dependent $\alpha_2$-semiaxis $2^{(D_C -1)/2}$. 
The surfaces are plotted in Fig.~\ref{fig:statespace}. 
In the figure, the ellipse that extends outside the state space and separates region A and B is the surface where $D_C = 0$. 
Inside that ellipse are thin dashed lines indicating $D_C = -1, -2, \ldots$, 
and outside are similar lines indicating $D_C = 1, 2, \ldots$.
The two other curves relate to symmetric extension which we will deal with next.

\subsection{Symmetric extension for cross sections}

If condition \eqref{eq:rankoneZsufcondsum} is satisfied, 
the corresponding state has a symmetric extension. 
Since the conditions \eqref{eq:rank2Zposconstraint1sum} and \eqref{eq:rank2Zposconstraint2sum} only fill the convex hull of this body, 
we will only need to consider the body described by inequality \eqref{eq:rankoneZsufcondsum} and its convex hull here. 

Unlike the surfaces for constant $D_C$, 
the surface of the set of extendible states is dependent on $\alpha_3$. 
In comparing symmetric extension to the $D_C$ surfaces, 
we will be particularly interested in three cross sections 
through the symmetric extension surface. 
One is through the center of the tetrahedron that defines the state space, 
where $\alpha_3 = 0$ ($p_x = p_y$). 
The two others are the two faces of the tetrahedron where
$\alpha_3 = \pm (1 - \alpha_1)/\sqrt{2}$ ($p_y = 0$ and $p_x = 0$).
% REMOVED IN THE FINAL STAGES. This part is not necessary.
% For states in one of these three cross sections, 
% the $\alpha_3$ coordinate (proportional to $p_x - p_y$) has a simple behavior 
% under B-steps.
% The ratio between $p_x$ and $p_y$ changes under B-steps as 
% \begin{equation}
% \frac{p_x^\text{out}}{p_y^\text{out}} = \frac{1}{2}\left(\frac{p_x}{p_y} + \frac{p_y}{p_x}\right) 
% \end{equation}
% in general, so in the case that $p_x = p_y$ they are still equal after a B-step.
% If $p_y = 0$, it will remain so throughout the B-steps, 
% as is easily verified from \eqref{eq:chau-recursion-y}.
% From this we also see that if $p_x = 0$, it is turned into $p_y = 0$ after one B-step.
% For states not in any of these cross sections, 
% the ratio $p_x/p_y$ is not conserved, but converges towards 1 from above.

For the cross section where $p_x = p_y$, we set $\alpha_3 = 0$ in 
Eq.~\eqref{eq:rankoneZsufcondsum} to get the equation
\begin{equation}
\label{eq:sixstatesymext}
  \frac{\alpha_2^2}{4} \left( 4\left(\alpha_1 - \frac{1}{2}\right)^2 + \alpha_2^2 -1)\right) = 0
\end{equation}
for the border. 
This tells us that any state with $\alpha_2 = 0$ (and at the same time $\alpha_3 = 0$) has symmetric extension, 
and they also happen to be separable. 
When $\alpha_2 \ne 0$, we get
\begin{equation}
\label{eq:sixstateellipse}
 4\left(\alpha_1 - \frac{1}{2}\right)^2 + \alpha_2^2  = 1
\end{equation}
which describes an ellipse with center in $(\alpha_1,\alpha_2) = (1/2,0)$, $\alpha_1$-semiaxis $1/2$, and $\alpha_2$-semiaxis $1$. 
In Fig.~\ref{fig:statespace} this is the solid curve separating regions C and D.

For the cases $p_y = 0$ and $p_x = 0$, we insert $\alpha_3 = \pm (1 - \alpha_1)/\sqrt{2}$ into Eq.~\eqref{eq:rankoneZsufcondsum}
to get 
\begin{equation}
  -\frac{1}{36}\left(\frac{9}{4}\left(\alpha_1-\frac{1}{3}\right)^2 + \frac{3}{2}\alpha_2^2 -1\right)^2 \geq 0,
\end{equation}
which simplifies to 
\begin{equation}
 \label{eq:surfaceellipse}
 \frac{9}{4}\left(\alpha_1-\frac{1}{3}\right)^2 + \frac{3}{2}\alpha_2^2 = 1.
\end{equation}
This describes another ellipse, with center in $(1/3,0)$, $\alpha_1$-semiaxis $2/3$, and $\alpha_2$-semiaxis $\sqrt{2/3}$. 
This is the solid line separating regions B and C in Fig.~\ref{fig:statespace}.

The outer ($\alpha_3 = 0$) symmetric extension curve (between C and D in Fig.~\ref{fig:statespace}) 
defines a border with no states with symmetric extension on the outside (for any $\alpha_3$). 
This is because if a state defined by $(\alpha_1,\alpha_2,\alpha_3)$ has a symmetric extension, 
so does the state defined by $(\alpha_1,\alpha_2,-\alpha_3)$ since the states are related by local unitaries. 
Then the convex combination $(\alpha_1,\alpha_2,0)$ would also have a symmetric extension. 
The inner symmetric extension curve [$\alpha_3 = \pm 1/\sqrt{2}(1 - \alpha_1)$, between B and C] 
is the border where all the states inside it has symmetric extension for all $\alpha_3$, 
since they can be obtained by mixing the states with symmetric extension on the surface of the state space.

\subsection{Distillability vs.~symmetric extension}

% Show connection between D_C and symmetric extension border
% - All states with D_C < 0 have symmetric extension (for alpha_3 = 0)
% - No states with D_C > 2 have symmetric extension 

We are now in a position to relate $D_C$ to symmetric extension.
From Fig.~\ref{fig:statespace}, it is evident that in most of the state space, 
the surface $D_C = 2$ lies strictly outside the outer border for symmetric extension (the line between C and D). 
Toward the point $(1,0)$, however, all the lines for constant $D_C$, symmetric extension border, and separability border converge. 
This is also the state toward which the sequence of states after the B-steps converges for the most relevant starting states (e.g.~all states  with $p_I \geq 0.5, p_z > 0$).
Even though this is a separable state, in any neighborhood around it there will be states without symmetric extension. 
%At this point the ellipses defining the symmetric extension border and $D_C = 2$ also have the same curvature. 
By inserting $D_C = 2$ in Eq.~\eqref{eq:chauconst}, the $D_C = 2$ border can be described by $\alpha_2^2 = 2 - 2 \alpha_1^2 =: f(\alpha_1)$. 
Similarly, the outer border for symmetric extension from Eq.~\eqref{eq:sixstateellipse} can be expressed as 
$\alpha_2^2 = 1 - 4(\alpha_1-1/2)^2 =: g(\alpha_1)$. 
Taking the difference, we get $\Delta(\alpha_2^2) = f(\alpha_1) - g(\alpha_1) = 2(\alpha_1-1)^2 \geq 0$, 
so the $D_C = 2$ surface is always outside the symmetric extension surface, 
except for the point $(1,0)$ where $D_C$ is not defined. 
Thus, no states for which $D_C \geq 2$ has a symmetric extension.
In a similar fashion, one can show that the Chau border $D_C = 0$ never is outside the inner symmetric extension border (between regions B and C) in the interval $\alpha_1 \in [0,1]$, 
which is the region where the Chau border is contained in the state space. 
They coincide at the points $(0,1/\sqrt{2})$ and $(1,0)$. 
Thus, any state with $D_C \leq 0$ has a symmetric extension.

To apply this to the six-state and BB84 QKD schemes, let us assume that Alice and Bob discard the
data specifying which bits come from which bases, meaning the error rates in the different bases are identical.
Only one possible state is consistent with the observed 
error rate $q$, namely, $q/2=p_x = p_y = p_z$. This immediately yields 
$q_\textrm{max} = (5-\sqrt{5})/10 \approx 27.64\%$ for $D_C = 0$.

From the BB84 measurements, only the error rates in the $x$ and $z$ basis can be observed, 
meaning the state is not completely determined, 
only that $q=q_x=q_z$ for $q_x := p_y + p_z$ and $q_z := p_x + p_y$.
If $q$ is below $1/2$, the possible eigenvalues are
$(1-2q, q, 0, q) + t(1,-1,1,-1)$ for $t \in [0,q/2]$.
When expressed in terms of $(\alpha_1,\alpha_2,\alpha_3)$, 
this becomes  $(1-2q,\sqrt{2}(1-3q),q\sqrt{2})+t(0,2\sqrt{2},-2\sqrt{2})$.
To determine if $D_C \leq 0$ for any of the possible states, 
we minimize $2^{D_C} = 2\alpha_2^2/(1-\alpha_1^2)$. 
This amounts to minimizing $|\alpha_2|$, 
since $\alpha_1$ is fixed by $q$, 
and it is obvious by inspection that $t=0$ gives the minimum. 
Solving $D_C = 0$, we find $q_\textrm{max} = 1/5$.

\subsection{Variations of B-steps}

% Equivalence of B-steps and CAD
In our analysis, we have used the B-steps introduced in \cite{gottesman03a} and inspired from 
classical advantage distillation (CAD) \cite{maurer93a}. 
CAD works on blocks of $N$ bits and Alice and Bob both announce the parities of all bits with the first bit. 
They keep the first bit if all the parities are equal, otherwise they discard the block.
%After the announcement the parities between any two bits is known, and given these parities 
Given the announced parities,
there are only two possible bit strings compatible with the announcements, namely, the correct string and the inverted string.
It is easy to see that when the block size for CAD is $N=2^n$, it is equivalent to $n$ successful rounds of B-steps 
(when the whole block is discarded if any of the B-steps fail).
To make sure that even for $N \ne 2^n$ CAD cannot break symmetric extension for any states were B-steps fail, we can compute the output state after CAD.
%CAD can be implemented on qubits by applying CNOTs from the first qubit to each of the other qubits, and then measure those other qubits in the computational basis to get the parities that are announced. 
The input state is $N$ copies of a Bell-diagonal state, which we think of as a maximally entangled state $\ket{\Phi^+}$ which has a probability $p_i$ for having suffered a $\sigma_i$ error on Bob's qubit.
The output qubit has a bit error (either $\sigma_x$ or $\sigma_y$) iff all the qubits in the block had a bit error and 
it has no bit error iff no qubit in the block had a bit error. 
The other bit error patterns are detected and the block is discarded.
The output qubit has a phase error ($\sigma_y$ or $\sigma_z$) if an odd number of input qubits had a phase error, 
and no phase error if an even number of input qubits had a phase error.
So the output qubit is error free if and only if an even number of input qubits had a $\sigma_z$ error and the rest were error free. 
The probability for this to happen given the state of the input qubits is 
$p_I^\text{CAD} = p_I^N + \binom{N}{2} p_I^{N-2} p_z^2 + \binom{N}{4} p_I^{N-4} p_z^4 + \cdots = \sum_{j=0}^{\lfloor N/2 \rfloor} \binom{N}{2j} p_I^{N-2j} p_z^{2j}$.
This is every second term in the expansion of $(p_I \pm p_Z)^N$, 
and by taking the average of the $\pm$ cases we get the terms we want, 
$p_I^\text{CAD} = \frac{1}{2}\left( (p_I + p_z)^N + (p_I - p_z)^N\right)$.
By making similar arguments, we get $p_z^\text{CAD}$ from the terms with an odd number of $\sigma_z$ errors, and $p_x^\text{CAD}$ and $p_y^\text{CAD}$ from the cases where there are $\sigma_x$ and $\sigma_y$ errors instead of $\mathbbm{1}$ and $\sigma_z$.
This gives the following generalization of Eqs.~\eqref{eq:chau-recursion}:
\begin{subequations}
\begin{align}
p_I^\text{CAD} &= \sum_{j=0}^{\lfloor N/2 \rfloor} \binom{N}{2j} p_I^{N-2j} p_z^{2j} \nonumber \\
&= \frac{1}{2}\left( (p_I + p_z)^N + (p_I - p_z)^N\right),\\ 
p_z^\text{CAD} &= \sum_{j=0}^{\lfloor (N-1)/2 \rfloor} \binom{N}{2j+1} p_I^{N-(2j+1)} p_z^{2j+1} \nonumber \\
&= \frac{1}{2}\left( (p_I + p_z)^N - (p_I - p_z)^N\right),\\ 
p_x^\text{CAD} &= \frac{1}{2}\left( (p_x + p_y)^N + (p_x - p_y)^N\right),\\ 
p_y^\text{CAD} &= \frac{1}{2}\left( (p_x + p_y)^N - (p_x - p_y)^N\right),
\end{align}
\end{subequations}
where the sum of these probabilities gives the probability for CAD to succeed.
From this, it follows directly that $D_C^\text{CAD} = N D_C$, 
so having the liberty to choose block sizes other than $2^n$ does not help if $D_C \leq 0$.

% Eqivalence of n-1 parities and CAD
Another observation is that the announcement of any $N-1$ independent parity bits on a block of $N$ bits is equivalent to performing CAD on a subset of $M\leq N$ of those bits. 
This can be seen simply by counting the number of possible strings.
On the block of $N$ bits, there are $2^N$ possible strings. 
Each announced parity bit halves this number, so after $N-1$ parity bits there are only two possible 
strings left. 
Any bits that are equal in the two strings are therefore completely revealed by the announcement.
The remaining $M$ bits are all different and the CAD on those $M$ bits is the same as announcing that we have one of those possible substrings.
Hence, nothing that generates a bit from $N$ bits by announcing $N-1$ independent parity bits can break the symmetric extension.

\section{Discussion}
\label{sec:discussion}

We have characterized the Bell-diagonal states that have a symmetric extension.
Using this we have shown that the failure of Chau's procedure to distill key from certain entangled states
can be understood in terms of failure to break a symmetric extension.
Also, some simple variations of the B-steps are shown to be equivalent with respect to distillability.
The natural question now is if any other modification of the procedure can distill key from these states.
Bae and Ac{\'i}n \cite{bae07a} attempted to improve the thresholds by 
adding noise in the beginning of the procedure, 
allowing coherent quantum operation on one side 
or measuring in a different basis, but without success.
Portmann~\cite{portmann05a} investigated BB84 with the bit error detection using random parities 
but could only prove security up to a QBER of 16.9\%.
We believe that the symmetric extension can be a useful tool for narrowing down  
which type of postprocessing, if any, can distill a secret key beyond current thresholds.

% Twirling can introduce symmetric extension
In the analysis, we have depended on the fact that the state after sifting can be considered to be Bell-diagonal, 
when one only considers the quantum bit error rate in the different bases.
While the state really is Bell-diagonal for a Pauli channel, 
it may be different in general and this would show up as correlations in the data where Alice and Bob measure in different bases. 
For other protocols, such as SARG04 \cite{scarani04a} and protocols based on spherical codes~\cite{renes04a}, 
the sifting works as a filter so the state will not be Bell diagonal even for a Pauli channel.
In these cases, the twirling procedure may actually turn a state without symmetric extension into one that has.
Any two-qubit pure state can be written in the Bell-basis
\begin{equation*}
 \ket{\psi} = \alpha_0 \ket{\Phi^+} + \alpha_1 \ket{\Psi^+} + \alpha_2 \ket{\Psi^-} + \alpha_3 \ket{\Phi^-}
\end{equation*}
and by choosing $|\alpha_j| = \sqrt{p_j}$, twirling will give a Bell-diagonal state with eigenvalues $p_j$.
If those $p_j$ are chosen such that the Bell-diagonal state has symmetric extension and the pure state is not a product state, 
the twirling will introduce a symmetric extension that was not there to begin with.
The most extreme example of this is when $\alpha_j = \exp(\mathrm{i} \pi j/2)$.  
Then the pure state is maximally entangled and since the correlations are in the wrong bases, the twirled state is maximally mixed.
A natural question is then what we can say about the symmetric extension for more general states, and this will be considered elsewhere~\cite{myhr08subb}.

\begin{acknowledgments}
The authors would like to thank Tobias Moroder for stimulating discussions.
This work was funded by the Research Council of Norway  (Project No.~166842/V30), the European Union through the IST Integrated Project SECOQC and the IST-FET Integrated Project QAP, Ontario Centres of Excellence, the NSERC Innovation Platform Quantum Works, and by the NSERC Discovery Grant.
J.M.R. acknowledges the support of the Alexander von
Humboldt Foundation and the Visiting Scholars Program at the
University of Queensland, where part of this work was carried out.
\end{acknowledgments}

% Use the ltxgrid command \onecolumngrid to go back to one column for the bibliography
\onecolumngrid

%\bibliography{qit_20080115,symextbell}
%\bibliographystyle{apsper} % Adds title to the bibliography

% --- Begin pasted bibliograpy ---

% --- End pasted bibliography

\end{document}